# Demographic Confounding Causes Extreme Instances of Lifestyle Politics on Facebook

Alexander Ruch[1], Yujia Zhang[2], Michael Macy[3]

[1] Spotify `aruch@spotify.com`; Cornell University, `amr442@cornell.edu`
[2] Cornell University, `yz658@cornell.edu`
[3] Cornell University, `mwm14@cornell.edu`

## Abstract

Lifestyle politics emerge when activities that have no substantive relevance to ideology become politically aligned and polarized. Homophily and social influence are able generate these fault lines on their own; however, social identities from demographics may serve as coordinating mechanisms through which lifestyle politics are mobilized are spread. Using a dataset of 137,661,886 observations from 299,327 Facebook interests aggregated across users of different racial/ethnic, education, age, gender, and income demographics, we find that the most extreme instances of lifestyle politics are those which are highly confounded by demographics such as race/ethnicity (e.g., Black artists and performers). After adjusting political alignment for demographic effects, lifestyle politics decreased by 27.36% toward the political "center" and demographically confounded interests were no longer among the most polarized interests. Instead, after demographic deconfounding, we found that the most liberal interests included electric cars, Planned Parenthood, and liberal satire while the most conservative interests included the Republican Party and conservative commentators. We validate our measures of political alignment and lifestyle politics using the General Social Survey and find similar demographic entanglements with lifestyle politics existed before social media such as Facebook were ubiquitous, giving us strong confidence that our results are not due to echo chambers or filter bubbles. Likewise, since demographic characteristics exist prior to ideological values, we argue that the demographic confounding we observe is causally responsible for the extreme instances of lifestyle politics that we find among the aggregated interests. We conclude our paper by relating our results to Simpson's paradox, cultural omnivorousness, and network autocorrelation.

## 1    Introduction

Political polarization and sectarianism, "the tendency to adopt a moralized identification with one political group and against another" (Finkel et al., 2020, p. 533), have steadily increased in Western democracies including the United States, Canada, New Zealand, and Switzerland since the 1970s (Boxell, Gentzkow, & Shapiro, 2020). Moreover, Americans' out-party hatred has exceeded their in-party love since 2012 – to where the former now better predicts voting behavior than the latter (Finkel et al., 2020). Identifying the cause of such discontents is complicated by the fact that polarization exists in many forms, including through party identification, political ideology, voting behavior, policy positions, and party sorting (Gentzkow, 2016). Despite these multiform definitions, little evidence shows that polarization is increasing by way of party identification, political ideology, or policy positions, but instead the strongest evidence of growing polarization comes from voting behavior (Bishop, 2009; Fiorina & Abrams, 2008; Gentzkow, 2016; Glaeser & Ward, 2005) and party sorting (Abramowitz & Saunders, 2008; Fiorina & Abrams, 2008). For example, Baldassarri and Bearman (2007) find that issue polarization changes social structure and partisans' interactional focus, leading them to believe that their in-groups are ideologically homogeneous when in fact they are heterogeneous. This issue partisanship then drives issue alignment among partisans, which leads to greater party sorting and coherence within parties on what their issue stances are (Baldassarri & Gelman, 2008).

Partisan polarization is not limited to explicitly political ideas and actions, however. Research by DellaPosta and colleagues (2015) finds that polarization is also increasing among lifestyles, leisure activities, consumption, aesthetic taste, and personal morality. They define instances of lifestyle politics such as these as "the curious formation of cultural enclaves among seemingly unrelated preferences… [on which] liberals and conservatives differ systematically on lifestyle dimensions that have no apparent substantive relevance to political ideology" (2015, p. 1475). Using 38 years of General Social Survey data, the authors also show how social dynamics of homophily (Mcpherson, Smith-Lovin, & Cook, 2001) and influence amplify lifestyle politics and how small biases stemming from individual level factors such as demographic backgrounds can act as coordinating mechanisms that activate and mobilize the generation of dependencies between individuals' attitudes, interests, and behavior and those of their network neighbors – a process they label network autocorrelation and liken to temporal and spatial autocorrelation (DellaPosta et al. 2015:1476). In other words, when similar sets of individuals interact, homophily and social influence can cause correlations to form between lifestyles and arbitrary but causally convicting attributes such as demographic characteristics and political ideologies as these network mechanisms drive group members to become increasingly aligned in their opinions (e.g., individuals across different social classes may not have inherent affinities for particular music styles because those genres have a special or unique class essence to them, instead music preferences may become entangled with social class as people from similar classes tend to spend more time with each other and listen to the same kinds of music together).

The claim that political and lifestyle issues may be arbitrarily aligned, entrenched, and polarizing between partisan groups may seem outlandish; however, Macy and colleagues (2019) used a multi-world experiment to show how path-dependent opinion cascades among liberals and conservatives can generate these exact outcomes. Using ten websites for measuring the opinions of partisans on different issues, where eight of websites involved identical intervention effects of social influence through which users could see how previous users evaluated each issue and where in two control websites no social influence existed and users could not see prior users' evaluations, the eventual partisan alignment of each issue was strongly correlated with the partisan alignment of the initial user to have evaluated the issue. In other words, in worlds in which a particular issue was first seen by liberals, liberals' decisions to approve of the issue caused liberals who visited the website later to also approve of the issue more often while causing conservatives to disapprove of the issue more often – and analogously for issues that were first evaluated by conservatives. The path dependency discovered also drove the same issue to be highly liberal in some worlds while driving the exact same issue to be highly conservative in other worlds – again, depending on the partisan alignment of the user who first evaluated the issue. This outcome is similar to that shown by Salganik et al. (2006), who found that evaluations of cultural artifacts such as music are strongly swayed through path dependent processes whereby songs that are positively evaluated first garner further positive reviews as individuals who confront a collection of unfamiliar content heavily defer to the previous evaluations of others to anchor their own opinions.

Lifestyle politics' presence extend far beyond opinion and music websites in ways that foreshadow troubling effects for Western democracies similar to those noted by Finkel and colleagues (2020). Liberals and conservatives, while similarly interested in science, do not consume the same science (F. Shi, Shi, Dokshin, Evans, & Macy, 2017). An analysis of network data representing millions of online book co-purchases shows that liberals prefer basic science over applied science, conservatives' preferred approach to science. The authors also found that "red" science books tended to be narrowly focused and located peripherally in the network, whereas "blue" science books spanned a substantially wider range of perspectives while also being more centrally located in the network. Such evidence could explain why partisans tend to talk past each other while discussing scientific matters from different frames. In another study, Shi and colleagues (2017) use Twitter co-following to infer political alignments and polarization in a sample of music, movie, hobby, sport, vehicle, food and drink, technology, university, religion, and business accounts. Their analysis finds strong evidence of lifestyle politics on social media as well, with stereotypes such as "Tesla liberals" and "bird hunting conservatives" having empirical validity. This network evidence is supported by survey evidence finding liberals and conservatives watch substantively different television (Blakley et al. 2020) and social media and internet browsing studies that reveal how partisans search for and subscribe to widely different news sources (Bakshy, Messing, & Adamic, 2015; Flaxman et al., 2016). Lastly, DellaPosta (2020) uses 44 years of nationally-representative panel data to argue that polarization may be growing through a trend of belief consolidation, whereby the presence of bipartisan alignments on a large number of topics is precipitously decreasing while at the same time partisan alignments among remaining topics are becoming increasingly clustered and cohesive to form "packages" of beliefs (e.g., religious beliefs becoming coupled with attitudes toward climate change, which then shape opinions on taxation).

Together, these findings on the evolution of lifestyle politics are concerning as they paint a picture of an America in which liberals and conservatives not only vote in starkly different ways based on their strong animosity toward one another, but they also understand and do science differently and are embedded in progressively different and divided ways of living – from what hobbies they have to television they watch at night to vehicles they drive down to what they eat and drink. This point may seem like an overstatement, but before the 2020 US Presidential Election, the *New York Times* (https://nyti.ms/3e2UpVc) published a quiz called "Can You Tell a 'Trump' Fridge from a 'Biden' Fridge?" where website viewers saw different pictures of open refrigerators and had to guess if the owners supported Trump or Biden based on what was in it. On average, after amassing 4.3 million votes on the photos, viewers guessed correctly only 52% of the time, leading one to believe that there are few partisan foods; however, select food were highly partisan at around 90%, including Kentucky Fried Chicken, Mountain Dew, and Krispy

Kreme doughnuts for Trump supporters and almond milk and flavored water for Biden supporters.

Overall, these points seem to provide clear evidence of how partisans in the United States see the world through different lenses and that they actively and passively – though implicit processes like network autocorrelation – carve out distinct spaces in it to live. What if, however, our interpretation of these partisan differences in lifestyles is misguided? What if, on the other hand, interests become polarized through a lifestyle politics process that is not driven directly by the actions of individuals from particular political parties but instead is driven by individuals from particular demographic groups that also happen to lean a particular way on the political ideology spectrum? For example, using the focal cultural artifact of DellaPosta et al. (2015) in their paper on lifestyle politics, lattes, if liberals drink lattes, is that happenstance because of who is liberal (e.g., mostly white, educated, high SES individuals) or is it because there exists an elective affinity between liberal ideology and lattes (e.g., "liberals hate America, and lattes are Italian")? DellaPosta and colleagues (2015) found that within-individual attributes such as demographic characteristics can act as coordinating biases through which such network autocorrelations are mobilized – leading demographic characteristics to become confounded with lifestyle politics in ways that can appear empirically convincing at the aggregate-level but which are actually the result of arbitrary alignments (e.g., social influence and homophily lead people with similar lifestyles to interact more and to shape each other's interests, often without any awareness of such influence and the resulting associations that diffuse between an interest in an activity and a group affiliation (Goldberg & Stein, 2018)). Such outcomes may also be caused by Simpson's paradox, where aggregate-level trends appear to lean in one direction but when analyzed at the subgroup level those trends disappear or even reverse for the subgroups.

To test this hypothesis of demographic confounding of partisan-interest alignments within lifestyle politics, we used a large network-based dataset representing 137,661,886 observations of 299,327 interests aggregated across Facebook users with different racial/ethnic, education, age, gender, and income demographics. We find substantial aggregate-level differences in partisan alignment across the 14,579 politically-relevant interests in the dataset. Topics on immigration and the Republican party are among the most conservatively-leaning interests at the aggregate-level; however, among the most liberal-leaning interests at the aggregate-level are Black and African American artists and performers – such that the aggregated measure of political alignment finds some of these interests to be 97.9% "liberal." After adjusting the alignment measure to deconfound for demographics, we find top-conservative interests focus on Republican themes and pundits while top-liberal interests now include things like electric cars, Planned Parenthood, and liberal satire. Moreover, the overall distribution of political alignment across interests shifted significantly more toward the center after adjusting for the effect of demographic-ideological entanglement as did interest topic distributions. These findings suggest that the rising prevalence of lifestyle politics we observe in the US may be caused by a tighter coupling between demographics and partisanship[1] versus growing affiliations between ideologies and different interests and activities, as demographics such as race/ethnicity, age, and gender are all causally prior to and predictive of political ideology.

---

[1] https://www.pewresearch.org/politics/2018/03/20/1-trends-in-party-affiliation-among-demographic-groups/

## 2    Data and Methods

### 2.1    Data and Sampling

In the spring of 2017, before Facebook restricted much of its API access[2] following the Cambridge Analytica scandal[3], we collected data on Facebook interests and the ideological and demographic characteristics of users who follow those interests using Facebook's Marketing API [4] by iteratively requesting aggregate information on the total number of US Facebook users who may be interested in a wide variety of interests intersected across multiple permutations of target users' ideological and demographic attributes[5]. For example, we not only searched for how many Facebook users may be reached by advertisements targeting individuals interested in musicians who happen to be liberal, but we also searched for how many users may like such interests who are liberal *and* Asian, liberal *and* Black, liberal *and* Hispanic, and so forth – extending this logic to include all possible intersections between each ideological category and all searchable demographic subgroups for race/ethnicity, education, age, gender, and income categories. Ideology in this context was coded by Facebook as leaning very liberal, somewhat liberal, moderate, somewhat conservative, or very conservative. The race/ethnicity category included three subgroups for Asian, Black, and Hispanic Facebook users. Education was binarized into subgroups of users who completed college and graduate, professional, or medical training and users who did not complete college or advanced educational training. Age was split into four bins: 13-21, 22-37, 38-57, and 57-100. We binarized gender as male or female. Lastly, we divided the income category into four subgroups: $30,000-$40,000, $40,000-$50,000, $50,000-$75,000, and $75,000+.

Using this sampling process, we collected aggregated reach data from 137,661,886 permutations across 299,327 interests and demographics. From these intersectional data, we created a dataset of interests with counts of users who like interests across ideological and demographic categories and dropped interests for which no ideological data existed, generating a dataset of 14,579 politically-relevant interests. This dataset includes the main set of observations we analyze later in the paper; however, while sampling, we also collected information on which interests are affiliated through users who co-follow any given pair of interests. We will use this data later in the paper to construct a network of interests in our main dataset, as the affiliation data comprise 6,130,012 edges between 14,300 interest nodes in the largest component.

We processed data in two more ways. First, we created columns for "liberal" and "conservative" followers by combining the reach values of "very" and "somewhat" liberal and conservative users, respectively. Secondly, as Facebook does not classify individuals as being White[6], to

---

estimate the number of White users who liked each interest we subtracted from the total reach of interests the number of Hispanic, Asian, and Black users who followed interests. This process generated a number of instances in which the estimated White reach of an interest was negative, indicating that the number of users who may belong to different demographic subgroups is overestimated. A Pew Research Center report[7] found that roughly 12% of Blacks and Hispanics identify as "multi-racial." Using this information, we down-weighted the counts of Black and Hispanic followers of each interest by 12% to address the potential double-counting issue, which led our estimates to return sensible non-negative counts of Whites.

Even with such adjustments, counts for race/ethnicity and other demographic categories may still have some error. For example, a Pew Research Center report[8] on Facebook's classification of users' demographics found up to 27% of people disagreed with how they were labeled in at least one category. Also, for any interest, if the aggregate count of individuals with a particular intersection of demographic characteristics is below 20, the Facebook Marketing API would censor the reach of this request to be capped at minimum of 20 to protect user anonymity. However, because the analyses in this study focus on larger intersections of ideological and demographic characteristics, these limits on censored data were rarely encountered. Also, given that Facebook earns 98% of its revenue from such advertisement tools[9], the company is motivated to ensure these classifications are accurate when applied in the context of interests at the aggregate-level, so we can be confident that the high-level observations we study among ideological and demographic trends in interests can improve our understanding dynamics involved in the spread of lifestyle politics.

## 2.2  Political Alignment and Demographic Deconfounding

Using interests' binarized liberal and conservative reach counts described in the section above, we developed a probabilistic estimate of interests' political alignment. For each interest, we subtracted 0.5 from the fraction of conservative followers among interests' liberal and conservative followers – assuming an even split in following between liberals and conservatives as a neural baseline value. We then multiplied this value by 2 to normalize political alignment between -1 (completely liberal) and +1 (completely conservative). Mathematically, these steps can be represented as follows:

$$Prob(I) \approx \frac{|C \cap I|}{|C \cap I| + |L \cap I|}$$

$$\pi_C(I) = 2(Prob(Conservative|I) - 0.5),$$

where $|C \cap I|$ denotes the number of interest $I$'s followers that are conservative and where $|L \cap I|$ is the analogous number for liberals. Since the denominator of the conservatively-oriented political alignment metric, $\pi_C(I)$, includes only the marginal following counts of conservatives and liberals and no other variables, it is symmetric to liberally-oriented political alignment. For

---

simplicity, we focused on the conservatively-oriented political alignment for the remainder of the analyses.

Because the aggregate-level political alignment for interests may be biased by Simpson's paradox, such as when the majority of an interest's followers are not necessarily liberal but instead belong to a demographic group that happens to be highly liberal, we developed a probabilistic approach to adjust interests' political alignment for various demographic confounding effects. Specifically, for each interest, we compared the political division within each demographic subgroup with the mean political alignment of each subgroup and calculated a weighted average of the differences. We can mathematically formulate this process as follows.

Let $\{D_i\}$ indicate the set of demographic subgroups comprising a particular demographic category. In the case of race/ethnicity, this would include Asian, Black, Hispanic, and White subgroups. We can then represent political division within each subgroup as $P\left(C|D_i \cap I\right)$, the level of conservative leaning among members of subgroup $D_i$ who follow interest $I$, and derive this value as follows:

$$P\left(C|D_i \cap I\right) = \frac{|C \cap D_i \cap I|}{|C \cap D_i \cap I| + |L \cap D_i \cap I|},$$

which is structurally similar to the unweighted aggregate-level measure of political alignment. It is important to note that because some intersections between political and demographic subgroups had no reach (0 followers), we could not use the marginal $|D_i \cap I|$ value in the denominator since it led to asymmetric values for conservative and liberal measures and induced uncertainty into the value's interpretation. For this same reason we also could not construct adjustment models for sets of intersecting demographics (e.g., race/ethnicity by gender), as the parametric space for the set of cross-comparisons to be made quickly exploded through the curse of dimensionality[10].

After calculating political division within each demographic subgroup, we represented the mean political alignment of each subgroup as $P\left(C|D_i\right)$, an overall feature of demographic subgroup $D_i$ that is independent of any specific interest $I$, and estimated such values as follows:

$$P\left(C|D_i\right) \approx \sum_{I_j \,\in\, all\ interests} \frac{|C \cap D_i \cap I_j|}{|C \cap D_i \cap I_j| + |L \cap D_i \cap I_j|}$$

Lastly, we take a weighted average of the differences between the subgroup political divisions and mean subgroup political alignment to produce our final estimate of demographically deconfounded political alignment for each interest. These weights indicate the impact of subgroups' presence in interest $I$. We can represent this factor as $P\left(D_i|I\right)$, the fraction of each group among all followers of $I$, which sums to 1 over all $\{D_i\}$. From this, we derive demographically deconfounded political alignment values as follows:

$$\mu_C\left(I; \{D_i\}_{i=1}^n\right) = \sum_{i=1}^n \left[P(C|D_i \cap I) - P(C|D_i)\right] \cdot P(D_i|I).$$

---

[10] The full parameter space to adjust interests' political alignment for all demographic subgroups collectively would include 512 (= 2*4*2*4*2*4) combinations, many of which would be empty and lead to further data sparsity issues.

We check that this expression is bounded between -1 and +1, as it is a weighted sum of probability differences that fall between -1 and +1. In fact, we may think of the previous measure of $\pi_C(I)$ as a degenerate case of the equation above where there is only one demographic group (all followers) and the group-level political division is assumed to be 50/50. Thus, this new measure provides a more general approach that encompasses the unweighted estimate and there is no logical conflict between the two equations. To measure the overall magnitude of change between the unweighted and weighted (deconfounded) political alignment distributions and to test whether these shifts are statistically significant, we used the `SciPy` (Virtanen et al., 2020) implementations Jensen-Shannon distance[11] and two-sample Kolmogorov-Smirnov statistic[12], respectively.

## 2.3 Topic-level Distribution Shifts

Studying how political alignment changes after demographic deconfounding over 14,579 interests collectively enables us to study the global trends in lifestyle politics for each demographic category; however, to better understand how demographic categories entwine with lifestyle politics among topical sets of interests, as Facebook provided no high-level schema for grouping interests into topics such as "music" or "sports" interests, we implemented two clustering methods. First, using information on how interests are affiliated through co-following, we created a graph of interests and used `graph-tool`, a high-performance Python library for graph modeling (Peixoto, 2014b), to perform high-resolution hierarchical stochastic block modeling (HSBM; Peixoto, 2014a). To account for the fact that some interests have stronger affiliation than others, the HSBM weighted edges across interest nodes by the number of people who co-follow each pair of interest nodes. From the hierarchical levels of interest topics discovered by the HSBM, we could select a level for which the model identified a set of topics with an appropriate range for analysis (e.g., between 15 and 30 topics of interests).

Because Facebook affiliation information alone may generate noisy clusters, as co-following may be influenced by filter bubbles and echo chambers more than thematic information about the topics themselves (Boutyline & Willer, 2017; Flaxman et al., 2016; Quattrociocchi, Scala, & Sunstein, 2016), we used a second clustering method using the Wikipedia knowledge graph and Latent Dirichlet Allocation (LDA) topic modeling. Specifically, we used the Wikipedia API to extract the set of categories and subcategories to which each interest belonged, selecting the most similar Wikipedia page for each interest when direct matches were not available. Some cases (827/14,579, 5.7%), however, had no matching Wikipedia pages and were excluded from topical analysis. From the remaining data, we removed uninformative (e.g., keywords on editing and archiving histories) and noisy tags (e.g., specific years for events and births). Because Wikipedia – like Facebook – does not provide high-level thematic topics (e.g., "music"), after we preprocessed the categories and subcategories we then concatenated these tokens into arrays that became category "documents" for each interest. These "documents" were then fed into a `gensim` (Řehůřek, 2010) `LdaMulticore` LDA model that was optimized to minimize perplexity and maximize coherence (Newman, Lau, Grieser, & Baldwin, 2010; Wallach, Murray, Salakhutdinov, & Mimno, 2009). The resulting topics were also inspected by the research team to assure they made sense, after which topics were provided with short names (e.g., "Art") to help interpretation.

---

[11] https://scipy.github.io/devdocs/generated/scipy.spatial.distance.jensenshannon.html
[12] https://docs.scipy.org/doc/scipy/reference/generated/scipy.stats.ks_2samp.html

# 3    Results

## 3.1    Unadjusted Aggregate-level Political Alignment

The distribution of unadjusted political alignment scores for the 14,578 interests, aggregated over all demographic categories, is presented in **Figure 1**. On average, Facebook interests had a political alignment of -0.064 (SD = 0.396) and were approximately normally distributed with 95% of the distribution's mass between -0.856 and 0.728. To verify that the interests we analyze were in fact politically relevant, for each interest we also calculated the fraction of followers that were liberal or conservative out of the total number of interest followers. These distribution for these political relevance values are presented in **Figure 1** as well. All interests in our data had a political relevance value above 0.5 (i.e., at least half of each interest's followers were classified as liberal or conservative), and the mean political relevance for interests was 0.755 (SD = 0.057), indicating that most interests in our dataset were substantively quite "political" in the sense that partisans followed them in large numbers (versus followed by ideologically moderate or non-political users).

**Figure 1: Unweighted Aggregate Distributions of Political Alignment and Political Relevance**

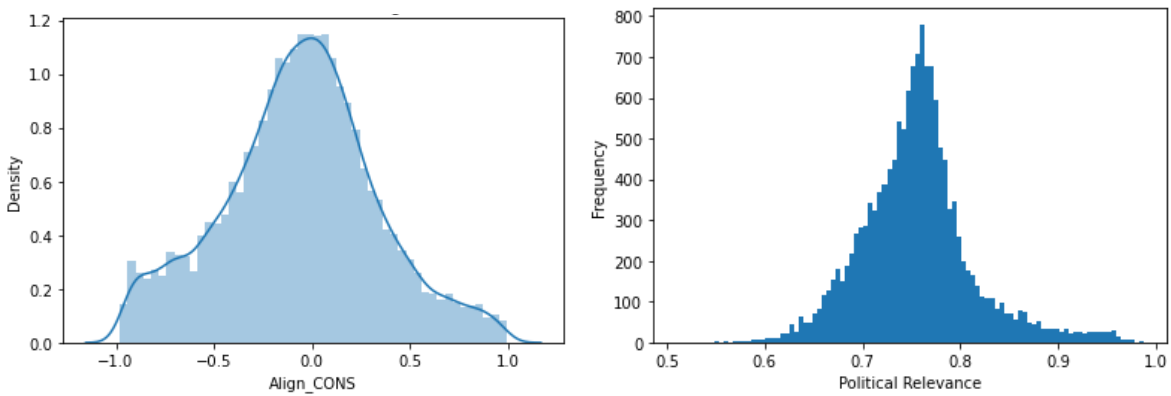

Notes: The unweighted aggregate distribution of conservatively-oriented political alignment has a mean of -0.064 and a standard deviation of 0.396. The unweighted aggregate distribution of political relevance has a mean of 0.755 and a standard deviation of 0.057.

When sorted in order of conservatively-oriented political alignment, the spread of lifestyle politics becomes immediately apparent. **Table 1** shows that whereas the most conservative interests in our dataset are in fact truly conservative – in the sense that preventing illegal immigration is a major conservative concern and that Brent Bozell and Rush Limbaugh are conservative commentators – the most liberal interests in our dataset at the aggregated-level are Black artists, cultural icons, and performers. In other words, whereas highly conservative interests appear to avoid strong lifestyle politics effects at the most extreme tail of the political alignment spectrum, highly "liberal" interests may be tightly entwined with lifestyle politics. Specifically, it is unlikely that these highly "liberal" interests are explicitly liberal at all. Instead, these interests likely only appear to be liberal at the aggregate-level because they are followed by many users who belong to demographic groups that also happen to be liberal (e.g., Black Americans in this case). In other words, demographics like race/ethnicity, education, age, gender, and income are likely confounding these aggregate-level estimates of interests' political alignments through the process of lifestyle politics – an effect we can directly measure using the equations derived in section 2.2.

**Table 1: Most Liberal/Conservative Interests in the Unweighted Aggregate Dataset**

| Liberal Interests | Political Alignment | Conservative Interests | Political Alignment |
|---|---|---|---|
| Dej Loaf | -0.979 | Stop Illegal Immigration | 0.991 |
| Jazmine Sullivan | -0.977 | Brent Bozell | 0.988 |
| Rasheeda | -0.976 | Rush Limbaugh | 0.986 |
| Lauren London | -0.976 | National Republican Senatorial | 0.986 |
| Chrisette Michele | -0.976 | American Patriots | 0.981 |

## 3.2   Global Deconfounding Results

To measure how entwined our set of Facebook interests are with lifestyle politics, we applied our demographic deconfounding process to our dataset to adjust the observed aggregate-level political alignment from the entanglements of race/ethnicity, education, age, gender, and income. **Figures 2–6** show how our original, unweighted, distribution of interests' political alignment changes after deconfounding for each of the aforementioned demographic effects. The five figures are presented in order of decreasing entanglement with lifestyle politics. As seen in **Figure 2**, race/ethnicity has the greatest association with lifestyle politics, as evident by the 0.1208 Jensen-Shannon distance between the adjustment weighted and original unweighted political alignment measures. This shift is also statistically significant according to a two-sample Kolmogorov-Smirnov test, which finds the maximum distance between the two distributions to be 0.2736 ($p < 1e\text{-}310$).

**Table 2** shows the most liberal and most conservative interests in the race/ethnicity deconfounded data to verify how our deconfounding weighting process removed the lifestyle politics effects of race/ethnicity. Now, the most liberal interests include electric concept cars, Planned Parenthood, and liberal comedians and politicians. By contrast, the most conservative interests revealed after deconfounding continue to be conservative pundits/commentators and states, Christian topics, and the largest GOP Facebook page in the world[13]. Race/ethnicity is the

---

only demographic category for which weighting leads Black artists and performers to no longer appear among the most "liberal" interests. For the results of the demographic categories that follow, the effects of lifestyle politics can still be observed in their weighted results by the fact that Black artists and performers remain among their most "liberal" interests despite there being a much narrower gap between liberals and conservatives along their weighted political alignment distributions. For this reason, we will not present any further examples of specific cases of top deconfounded liberal/conservative interests.

Deconfounding political alignment from the lifestyle politics effects of education, age, gender, and income demographics reveal statistically significant shifts from the original unweighted measures, albeit at a smaller magnitude compared to that of race/ethnicity. For example, as shown in **Figure 3**, deconfounding political alignment of educational subgroup effects generates a new distribution of weighted measures that have a Jensen-Shannon distance of 0.1065 and a Kolmogorov-Smirnov maximum distance of 0.2200 (p = 1.864e-310) from the original aggregated measures of political alignment. In **Figure 4**, the deconfounding results for our four age subgroups are similar, leading to weighted measures of political alignment that have a Jensen-Shannon distance of 0.1032 and a Kolmogorov-Smirnov maximum distance of 0.2103 (p = 2.407e-283) from the original unadjusted measures. Deconfounding for age (**Figure 5**) leads to weighted measures with a Jensen-Shannon distance of 0.097 and a Kolmogorov-Smirnov maximum distance of 0.1823 (p = 2.285e-212) from the original aggregated political alignment measures. Finally, **Figure 6** reveals how deconfounding for income subgroup effects leads to weighted measures that have a Jensen-Shannon distance of 0.095 and a Kolmogorov-Smirnov maximum distance of 0.165 (p = 4.100e-173) from the original unweighted political alignment measures.

Overall, these global deconfounding results demonstrate how adjusting the aggregate measures of political alignment for demographic effects significantly narrows the span of alignment inward to the political center while also correcting alignment values that previously appeared to be extremely "liberal" or "conservative" when observed in the aggregate. In other words, these results show that it is not the case that Black artists and performers are truly the most liberal interests in the sense in which we commonly think of "liberal" interests. Instead, these interests are simply overrepresented among the aggregate "liberal" interests due to their demographic confounding, as Black Americans tend to be liberal and if Black Americans tend to be more interested in Black artists and performers then this entanglement is what causes lifestyle politics to spread into such genres of entertainment. Note that causal reasoning is intentional here, as demographic characteristics like race/ethnicity, age, and gender exist before and are causally prior to political ideology; therefore, the only logical way to understand demographic entanglement with lifestyle politics is via paths like race/ethnicity shaping political alignment which then can become emphasized over race/ethnicity when interests are analyzed in the aggregate without consideration of the audiences that engage in those lifestyles. Again, these results are also analogous to those seen in instances of Simpson's paradox in the sense that at the aggregate-level many interests that appear to lean strongly "liberal" were actually only somewhat liberal or moderate after political alignment was adjusted for subgroup-specific trends.

**Figure 2: Political Alignment Distributions before/after Deconfounding for Race/Ethnicity**

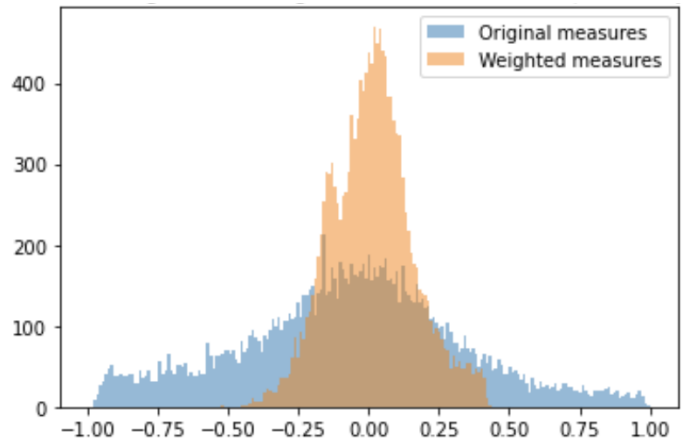

Notes: Jensen-Shannon distance 0.1208, two-sample Kolmogorov-Smirnov statistic 0.2736 (p ~ 0.0).

**Table 2: Most Liberal/Conservative Interests in the Race/Ethnicity Deconfounded Dataset**

| Liberal Interests | Political Alignment (%Δ) | Conservative Interests | Political Alignment (%Δ) |
|---|---|---|---|
| Nissan Pivo | -0.525 (-8.1%) | Greg Abbott | 0.434 (-52.8%) |
| Adam Schiff | -0.456 (-52.8%) | Sid Miller | 0.425 (-56.5%) |
| Planned Parenthood | -0.440 (-53.7%) | Republic of Texas | 0.425 (-53.5%) |
| John Oliver | -0.428 (-50.3%) | Christian Zionism | 0.424 (-54.7%) |
| Andy Borowitz | -0.424 (-52.9%) | Positively Republican! | 0.424 (-56.6%) |

**Figure 3: Political Alignment Distributions before/after Deconfounding for Education**

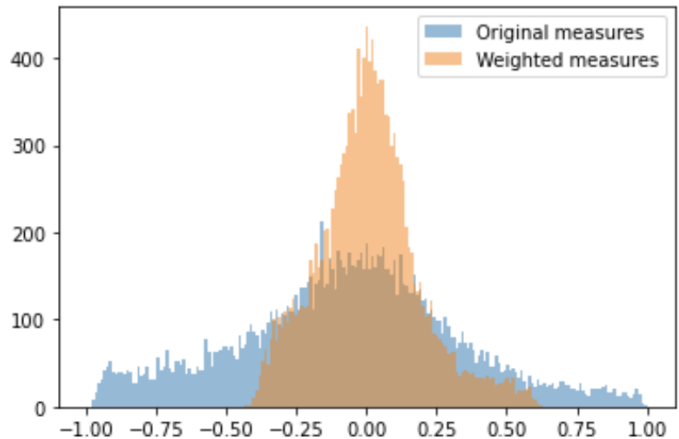

Notes: Jensen-Shannon distance 0.1065, two-sample Kolmogorov-Smirnov statistic 0.2200 (p ~ 0.0).

**Figure 4: Political Alignment Distributions before/after Deconfounding for Age**

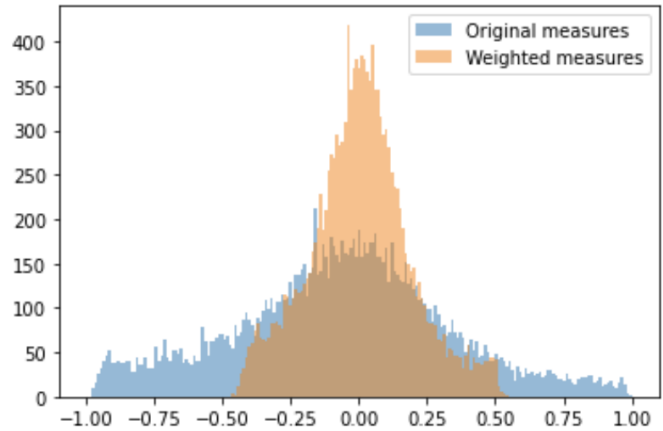

Notes: Jensen-Shannon distance 0.1032, two-sample Kolmogorov-Smirnov statistic 0.2103 (p ~ 0.0).

**Figure 5: Political Alignment Distributions before/after Deconfounding for Gender**

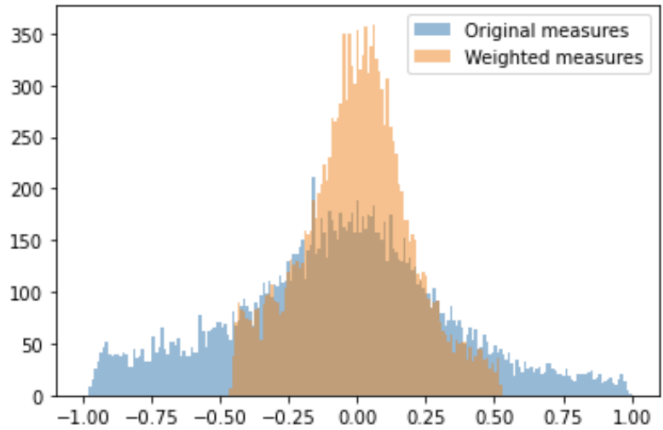

Notes: Jensen-Shannon distance 0.0969, two-sample Kolmogorov-Smirnov statistic 0.1822 (p ~ 0.0).

**Figure 6: Political Alignment Distributions before/after Deconfounding for Income**

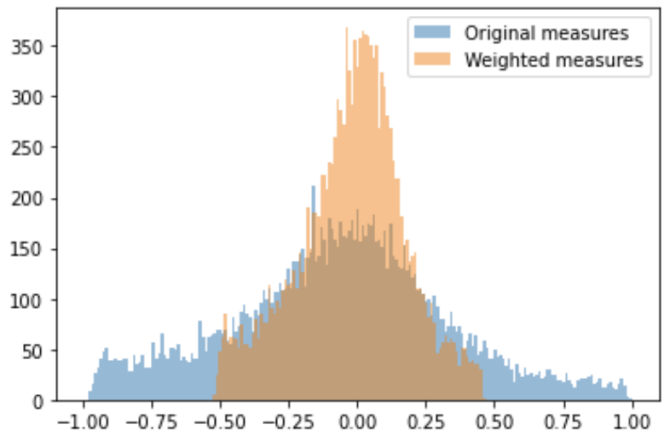

Notes: Jensen-Shannon distance 0.0952, two-sample Kolmogorov-Smirnov statistic 0.1646 (p ~ 0.0).

## 3.3 Topic-level Deconfounding Results

While evaluating global shifts in political alignment before and after demographic deconfounding informs our understanding of the magnitude to which each demographic category is entwined with lifestyle politics across politically-relevant interests as a whole, we also analyzed how alignments changed after deconfounding over sets of interests to understand more clearly how certain topics that liberal and conservative Facebook users follow may shift by different magnitudes. Analyzing topic-level changes in political alignment before and after adjusting for demographic effects can inform us of whether different demographic characteristics are especially entangled with certain lifestyle topics above and beyond their associations with other kinds of lifestyle topics. It may be the case, for example, that many interests relate to entertainment and that demographics are mostly entangled with lifestyle politics among entertainment topics but that little confounding associations exist between demographics and other topics that may appeal to a wider range of individuals across our full set of ideological and demographic groups, such as cars. Additionally, by analyzing topic-level shifts in alignment before and after deconfounding, we can also assess if some demographic characteristic is much more entwined with the lifestyle politics of a particular topic compared to other demographic characteristics that may be entwined with the lifestyle politics of that topic. For instance, maybe race/ethnicity is primarily driving the lifestyle politics we observe among music and adjusting the music topic's alignment for other demographics leads to little relative change.

To create these topics of interests, we first performed hierarchical stochastic block modeling on our interest network. **Figure 7** shows the interest network of 14,300 nodes and 6,130,012 edges is divided along liberal and conservative lines, with the top-left portion of the network leaning more liberal and bottom-right part of the network leaning more conservative. The hierarchical stochastic block model discovered 8 hierarchical levels in which nodes can be grouped into higher-ordered clusters of an increasing number of nodes. We chose to analyze the $5^{th}$ hierarchy of clustering, as this generated 18 clusters of nodes total[14]. Unfortunately, the topics generated from this block modeling approach suffered from two major problems. First, incoherent combinations of interests were often present (e.g., one block was comprised of interests related to children's clothes and the government). Second, since clusters found by the block model included liberal and conservative interests together in the same cluster, when we measure topic-level changes in political alignment after deconfounding we would not be able to tell whether liberal topics specifically shifted more than conservative topics.

---

[14] Cutting the hierarchy at level 4 would have generated 61 clusters of nodes, whereas cutting it at level 6 would have generated 5 clusters. We manually the inspected the clustering results of levels 4 – 6 and determined a cut at level 5 was optimal as the results from the level 4 cut were often too focused and the level 6 cut was too aggregated.

**Figure 7: Interest-to-Interest Network Projection with Political Alignment Annotations**

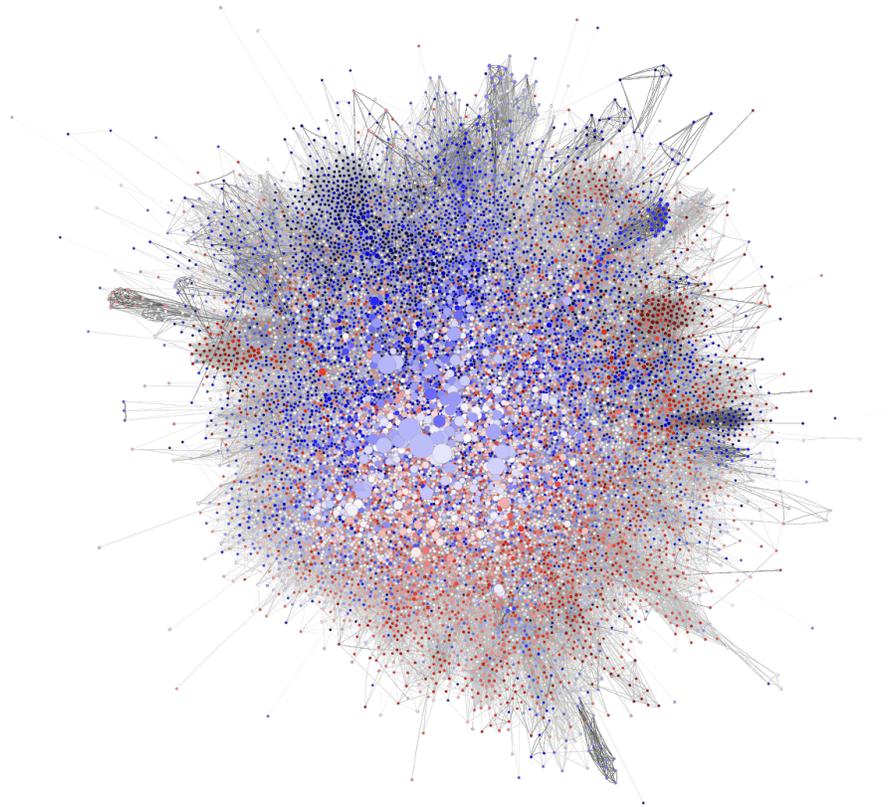

Note: The interest-to-interest network contains 14,300 nodes and 6,130,012 edges. Each node represents a Facebook interest. Nodes are annotated for size (larger nodes are followed by more users) and political alignment (dark blue nodes lean heavily liberal and dark red nodes lean heavily conservative). Edges represent co-following between two interests among the same set of users, where thicker edges indicate a larger number of co-followers.

Instead of using a purely network-driven approach to discovering topics of interests among our data, we used network and language data together to generate our topical sets of interests. By performing LDA topic modeling on category information available on 13,752 of our interests via Wikipedia's knowledge graph API, we were able to discover 31 liberal topics and 22 conservative topics[15], from which we could measure of how much political alignment shifted for the two parties after deconfounding across the five demographic categories. For example, in **Figure 8** we can see that after deconfounding (light blue/red points) both liberal and conservative topics had wide and statistically significant shifts from their unweighted aggregate-level values of political alignment (dark blue/red points) toward the neutral middle of the alignment scale. Liberal topics on average not only had wider post-deconfounding shifts versus conservative topics but they also shifted closer to the neutral middle ground compared to their conservative counterparts. Although these post-deconfounding topic-level shifts in alignment are substantial (e.g., the liberal "singers" topic shifted 0.4397), they are not wide enough to lead any topic to become statistically indifferent from the 0.0, the neutral value of alignment. Therefore, demographic categories such as race/ethnicity are not enough to fully explain away the partisan differences within these topics, and they persist in holding onto an inherent political lean that may be beyond the effects of lifestyle politics.

---

[15] See Appendix A1 and A2 for intertopic distance maps representing the diversity and clustering of these results.

The topic-level deconfounding results for demographic categories such as education, age, gender, and income are similar to those discussed for race/ethnicity, albeit they have narrower shifts post-deconfounding than those observed with race/ethnicity. For example, after deconfounding for the education effects (**Figure 9**), liberal topics continue to shift more toward a neutral alignment than conservative topics generally do. Post-deconfounding alignment shifts across both ideologies were statistically significant at the $p < 0.05$ level for all interest topics – as they were for race/ethnicity. **Figures 10 – 12** for age, gender, and income, respectively, show similar trends to those discussed here; however, across each of these respective demographic categories, the range of deconfounded shifts becomes narrower and narrower compared to race/ethnicity and education; however, despite these overall differences in magnitude, across topics we observe no instances in which any specific demographic is primarily responsible for driving the largest change in alignment after we apply our adjustments. In other words, the demographic effects of race/ethnicity, for example, tend to be well-distributed over interests from each of the identified topics and do not drive lifestyle politics substantially more among some sets of topics versus other topics any more than the between-topic differences we observe after deconfounding for any of the other demographic categories. These trends in how alignment distributions change after demographic deconfounding follow the same patterns in magnitude by demographic group as those revealed by Jensen-Shannon distances and Kolmogorov-Smirnov statistics from Section 3.2.

**Figure 8: Political Alignment Distributions before/after Deconfounding for Race/Ethnicity**

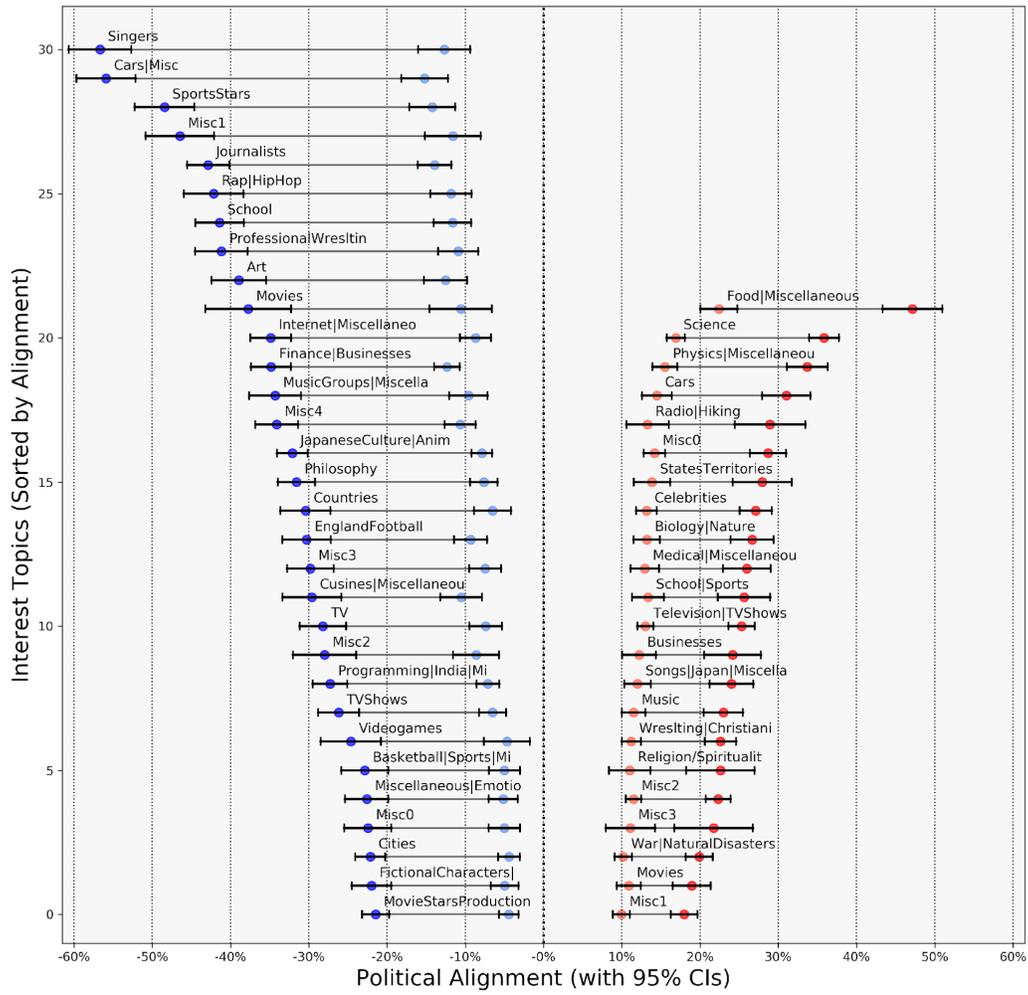

**Figure 9: Political Alignment Distributions before/after Deconfounding for Education**

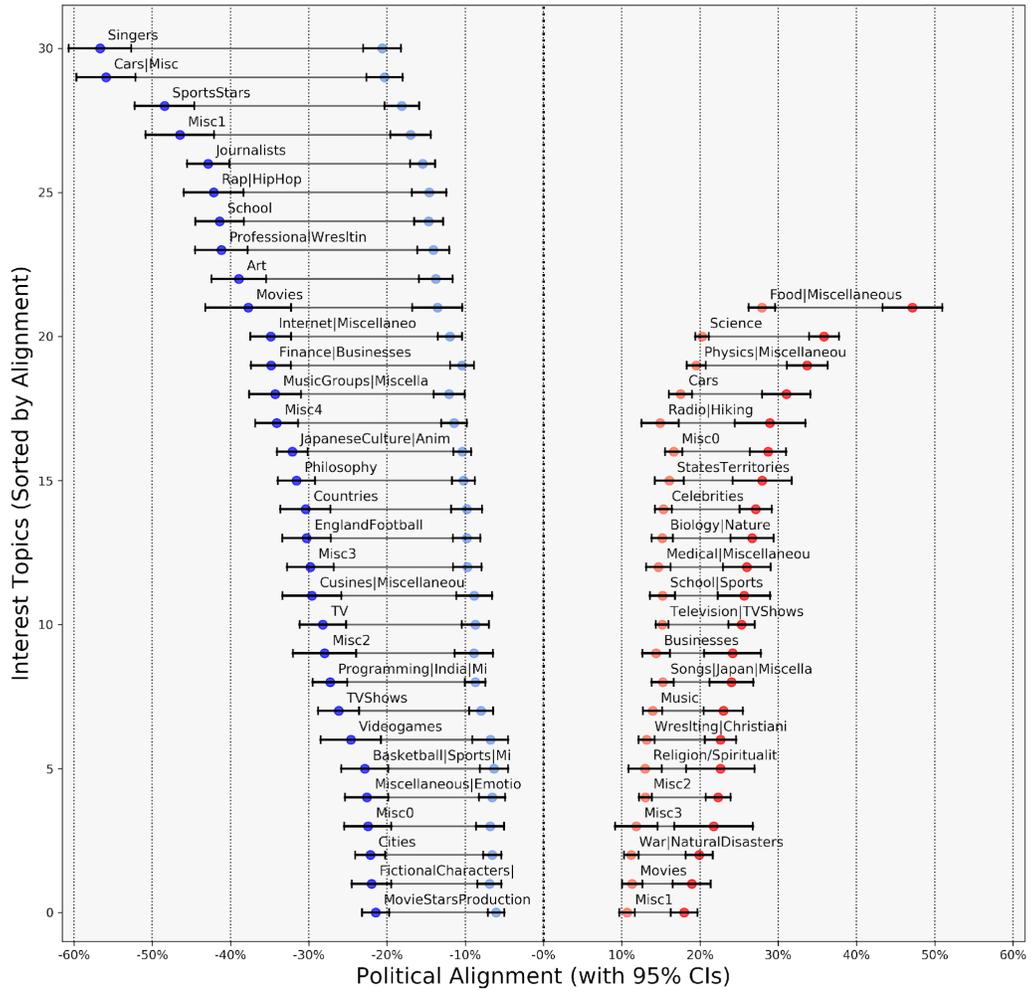

**Figure 10: Political Alignment Distributions before/after Deconfounding for Age**

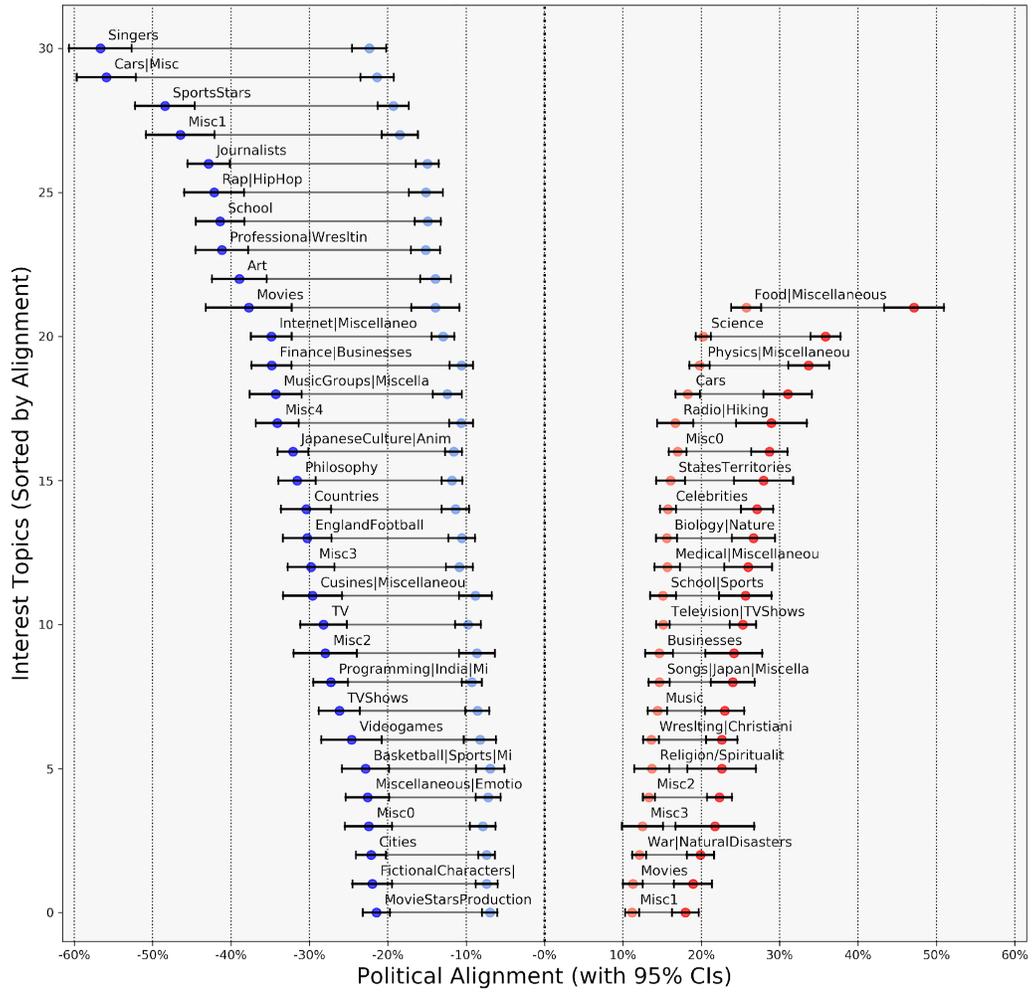

**Figure 11: Political Alignment Distributions before/after Deconfounding for Gender**

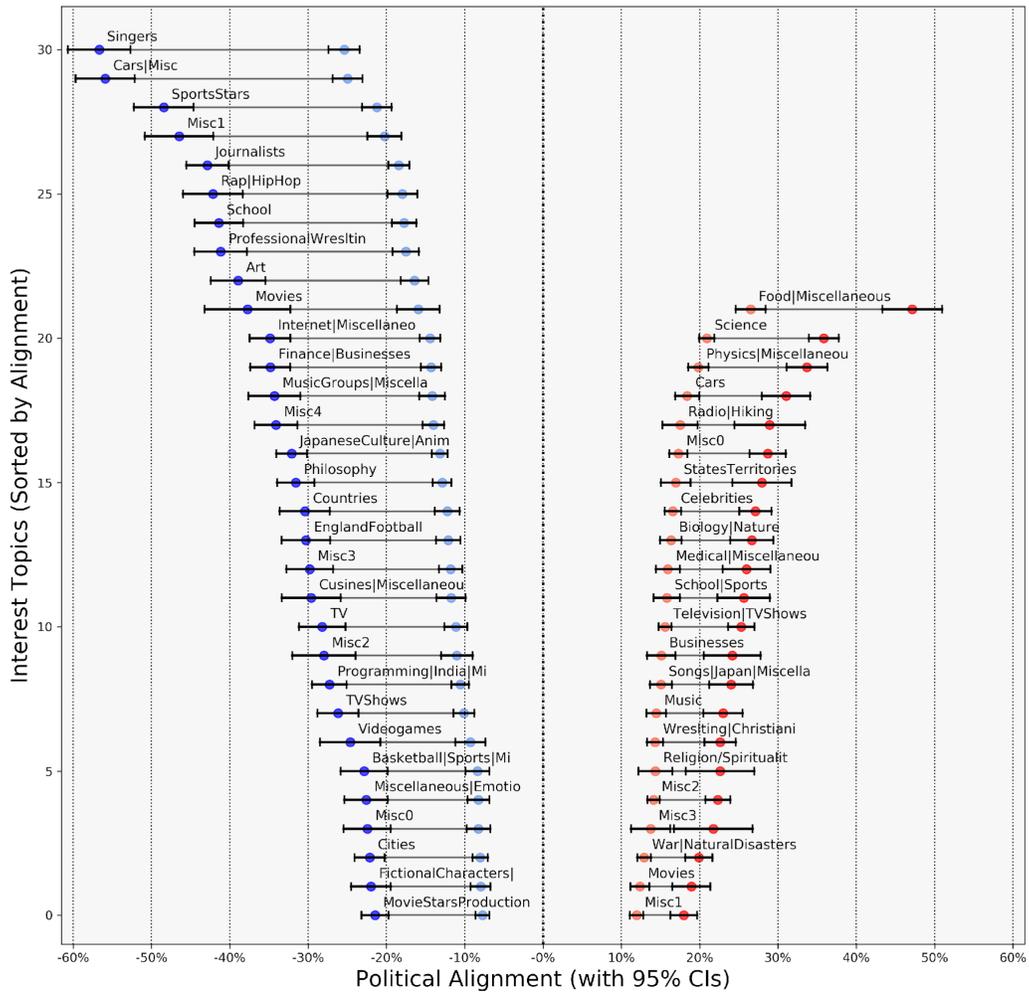

**Figure 12: Political Alignment Distributions before/after Deconfounding for Income**

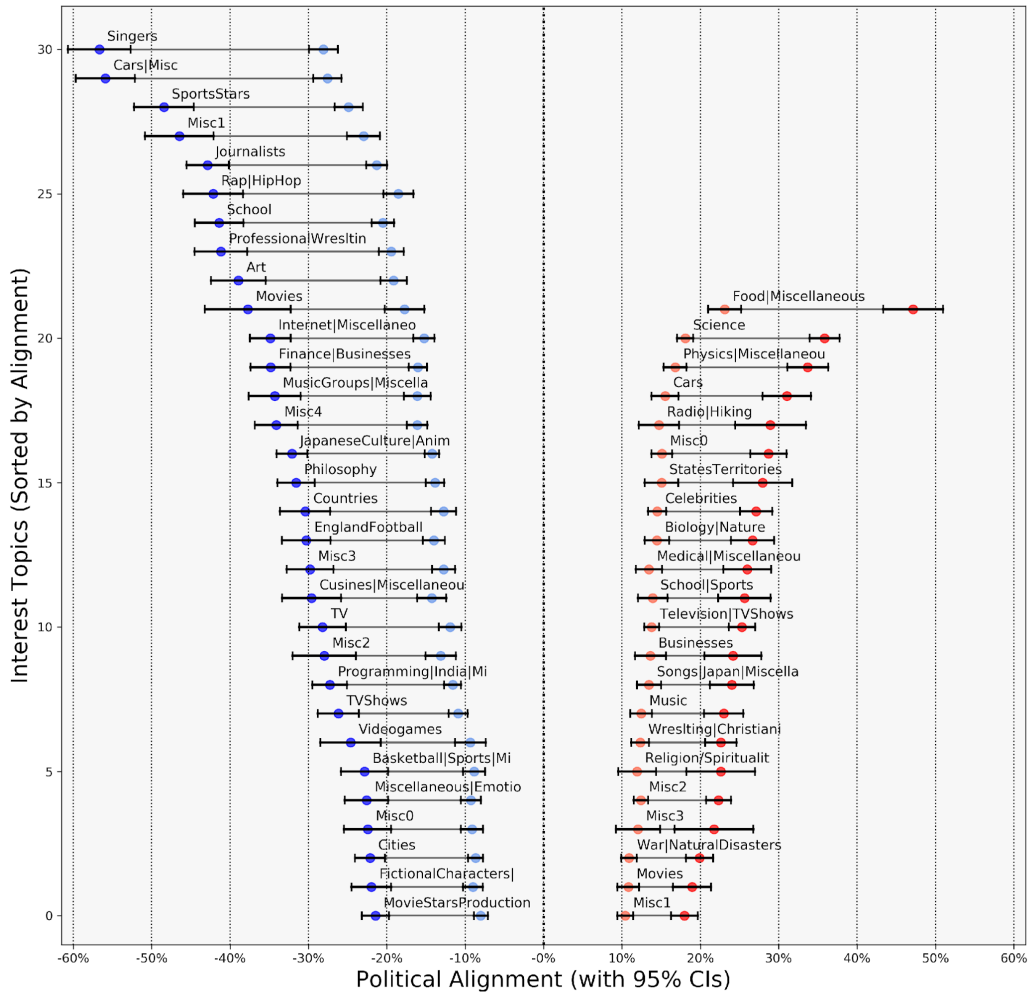

## 3.4  Measurement Validity Test

To test the validity of our political alignment measures for interests and whether interests' political alignment on Facebook is similar to the observed political alignment of such interests using other, possibly more internally valid methods, we used the data from the General Social Survey. The first validation test we performed was on whether interest in rap music[16] was politically aligned and if such an alignment would be attenuated or removed after controlling for respondents' race. In the bivariate case, political ideology (encoded on a 7-item Likert scale with higher values indicating stronger conservativism) had a linear association of 0.086 (p = 0.0012) with interest in rap music (encoded on a 5-item Likert scale with lower values indicating greater interest); however, when race is added to the model as a binary indicator variable (White vs. Black), then political ideology's linear association with interest in rap dropped to 0.069 and became statistically insignificant (p = 0.102) whereas the indicator variable for Whites became statistical significant (p = 3.64e-9) with a larger coefficient of 0.801. This result is analogous to what we observe when we deconfound our interests for race/ethnicity and discover that singers and other music topics become significantly less politically aligned toward liberalism (compare **Table 1** to **Table 2** and see **Figure 8**).

---

[16] The most recent General Social Survey in which interest in rap music was surveyed was 1993 (https://gssdataexplorer.norc.org/projects/88944/variables/1400/vshow). While it is true that interest in rap music may have become more diffuse across demographic subgroups beyond Blacks since 1993, the fact that the statistical tests described in the paper for these survey results align with what is observed in our demographic deconfounding results for Facebook interests gives evidence that the lifestyle politics effects we capture and measure are valid.

In our second validation test, we first regressed agreement with the statement "Immigration from Latin America should be increased[17]" (encoded on a 5-item Likert scale where higher values imply greater disagreement) against the aforementioned 5-item political ideology variable, resulting in a statistically significant association between the two variables of 0.053 (p = 0.0238). When we add age into the regression model, however, the linear association for ideology decreases to 0.045 and becomes statistically insignificant (p = 0.056) whereas the coefficient for age (encoded discretely) is statistically significant with a coefficient of 0.007 (p = 1.19e-4). This result also validates what we observed for the conservatively leaning interest "Stop Illegal Immigration" in **Table 1**, which had an aggregate-level alignment of 0.991 before deconfounding for age and which had a post-deconfounding alignment of 0.514. Together with the rap music interest results noted above, these two survey-based validation tests give strong evidence that the alignment and adjustment trends that we observe in our dataset of aggregate internet-based trace behavior are valid.

# 4    Conclusion

Social media is framed as a core theater in the culture wars; however, demographic deconfounding across categories such as race/ethnicity, education, age, gender, and income reveals the magnitude of partisan differences is actually significantly less than appears to be when observed in aggregate across a wide range of interests – following a pattern akin to Simpson's paradox in lifestyle politics. After adjusting interests' political alignments for their associations with these demographics, we also discovered a pattern of liberal topics of interests being more strongly entangled with lifestyle politics through these demographics than conservative interest topics, possibly due to a process of cultural omnivorousness in which people with elite socioeconomic characteristics such as higher education and income who have a high tolerance for multiculturalism, racial/ethnic diversity, and cultural inclusivity tend to engage with a wide variety of cultural interests and artifacts (Goldberg, 2011; Peterson, 1992; Warde, Wright, & Gayo-Cal, 2007). Our outcomes aligned with results from validation tests performed on the General Social Survey, including one from 1993, indicating that the demographic drivers of lifestyle politics we observe in interests are not likely to be caused by social media echo chambers or filter bubbles, as social media did not exist as ubiquitously then as it does now.

Because demographic characteristics such as race/ethnicity, age, and gender are causally prior to ideological values like liberalism and conservatism, our results demonstrate how lifestyle politics we observe among many interests are causally spurred through demographic entanglement. In the case of Black and African American artists and performers appearing to be the most "liberal" cases among our topics, for example, the fact that adjusting the aggregate political alignments of these interests for the fact that Black Americans tend to be liberal, which then lead these interests to be far less liberal compared to other stereotypical liberal interests like electric vehicles and Planned Parenthood, shows how the lifestyle politics of these issues was generated through demographics and not through simple network autocorrelation processes such as homophily and social influence. That being said, it is likely the fact that network autocorrelation plays a strong role in determining why individuals who are non-White, highly education, and/or of higher income tend to lean liberal; however, here we observe that lifestyle politics of these musical cases are more so driven through paths that lead from race/ethnicity (Black) to ideology (liberal), after which the lifestyles in which individuals with those demographic backgrounds participate (music) come to be labeled as "liberal" when they are observed in aggregate without consideration of the subgroups that form the audience and participants of those interests and activities. To be pedantic, it cannot possibly be the case that the lifestyle politics we observe among these interests are driven through a relationship in which liberal ideology generates the racial/ethnic backgrounds of those interests' audiences.

---

[17] This variable comes from the 2000 General Social Survey
(https://gssdataexplorer.norc.org/projects/88944/variables/2304/vshow) and is closely related to Facebook interests that appear to lean heavily conservative when aggregated across all demographic categories.

DellaPosta and colleagues (2015) also observed through their simulations of lifestyle politics and the social dynamics that generate them that whereas homophily and social influence can create the lifestyle politics we observe among many ideas, values, and behaviors in absence of social identity factors like demographics, lifestyle politics are more effectively mobilized when they form around such identities. In our models, we also found that the most "liberal" interests tended to be those that were heavily followed by non-White and/or highly educated users, and so we found direct evidence to empirically support this hypothesis that lifestyle politics is more effectively mobilized around demographic characteristics. While this may be true among the most ideologically extreme cases of interests that are entwined with lifestyle politics, it could be the case as DellaPosta (2020) argues that interests that are more moderately affected by lifestyle politics are aligned with partisan groups through an evolving belief network whereby cross-cutting associations between interests among partisans are becoming less common over time. For instance, after we adjusted interests for demographics, one of most liberal interests was an electric car while one of the most conservative interests was the Republic of Texas. Perhaps the reason why liberals like electric cars is because Texas is not only conservative but is also an oil-producing state, and so since liberals oppose things that are conservative, they oppose gas-powered cars and came to support electric vehicles. This is similar to what Macy and colleagues (2019) found in their experiment on opinion cascades across different policies, where once a member of one ideological group claimed support for a policy that was previously neutral, that issue then became associated with that party and garnered even more support from individuals from that same party and attracting opposition from members of the other party. Because whoever came to support each issue was arbitrary based on the randomization they employed in their experiment, in some experimental worlds an issue that became highly liberal was highly conservative in other worlds. Put another way in the context of our study, perhaps if in another world Texas was liberal, we would have observed electric cars among the top conservative interests after we adjusted for demographic entanglement, given that our demographic adjustments cannot completely explain the variation in lifestyle politics we observe among interests.

Though our study cannot fully account for the presence of lifestyle politics on Facebook, we were able to show how at least 27.36% of the magnitude of lifestyle politics among interests is largely explained by the demographic characteristics of individuals who follow those interests. In doing so, our research empirically supports sociological and social psychological theories on lifestyle politics, including hypotheses on how demographics can serve as social identities around which lifestyle politics are mobilized. Because we were only able to analyze aggregate data and were not able to measure how individuals who belonged to intersecting sets of demographics were interested in different interests, future studies can test whether the evolving belief network hypothesis of DellaPosta (2020) is able to better explain the persistence of lifestyle politics within social media than our argument about demographic entanglements. Also, since our data are not temporal, future studies can test whether and how such entanglement is becoming stronger or weaker over time, given that the associations between ideology and demographics are becoming stronger over time while on the other hand the gaps between liberals and conservatives across their support for various values have been diminishing with both groups becoming more secular over time (Baldassarri & Park, 2020).

## Acknowledgements

This work was supported in part by NSF grant SES-1756822. The views and conclusions presented in this study are those of the author and should not be interpreted as necessarily representing the official policies or endorsements – either expressed or implied – of NSF, the U.S. Government, Cornell University, or Graphika Inc. The author wishes to thank Agrippa Kellum, James Zou, and Alex Siu for assistance with collecting Facebook interest data, evaluating topic model results, and validating our mathematical models, respectively. The author also wishes to express gratitude to Cornell University's Social Dynamics Laboratory and Cornell University's class on "Culture Wars and Polarization" for comments and feedback on parts of this project.

# Appendix

## A1   LDA Intertopic Distance Map: Conservative Topics (*n* = 22)

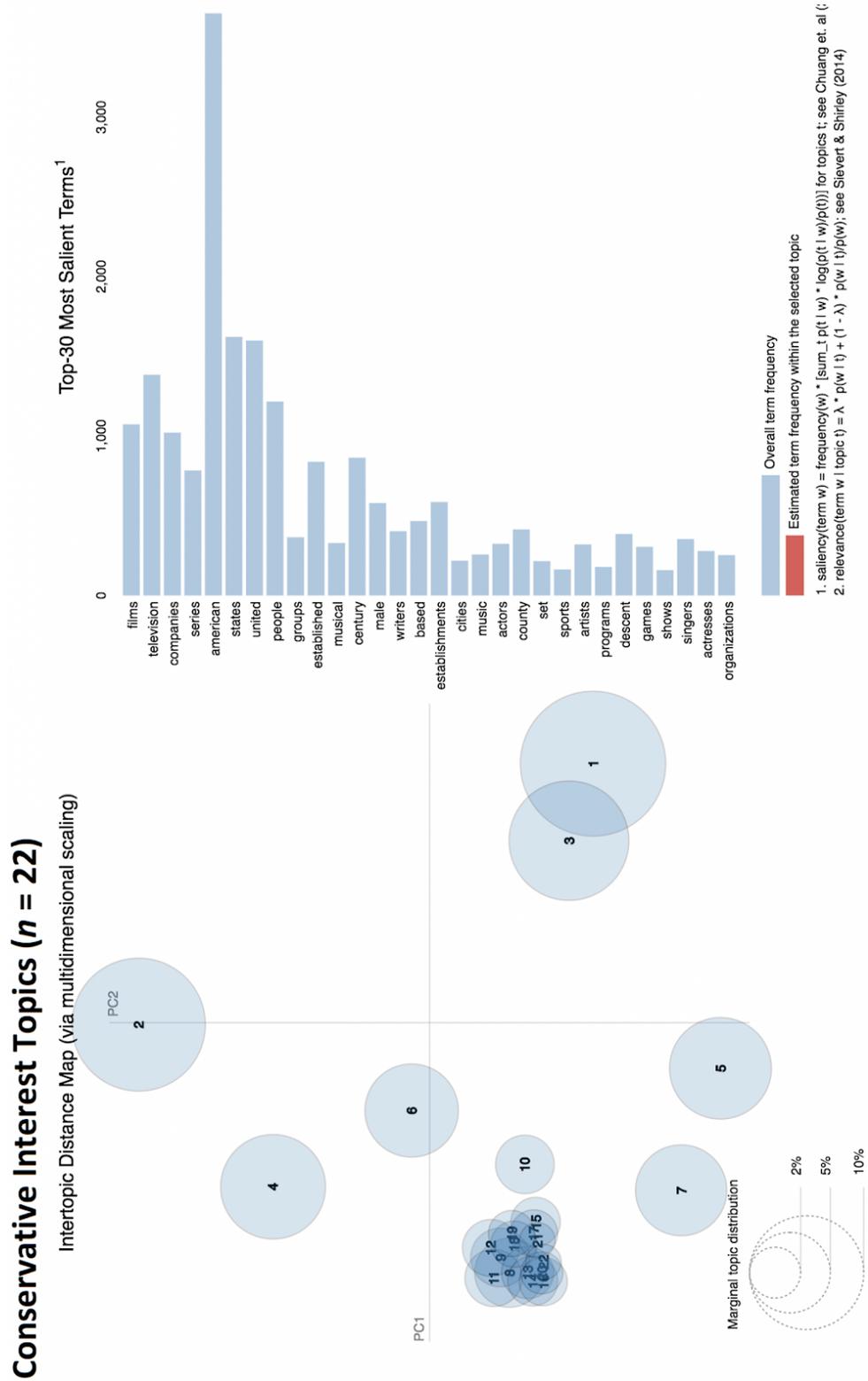

## A2 LDA Intertopic Distance Map: Liberal Topics (*n* = 31)

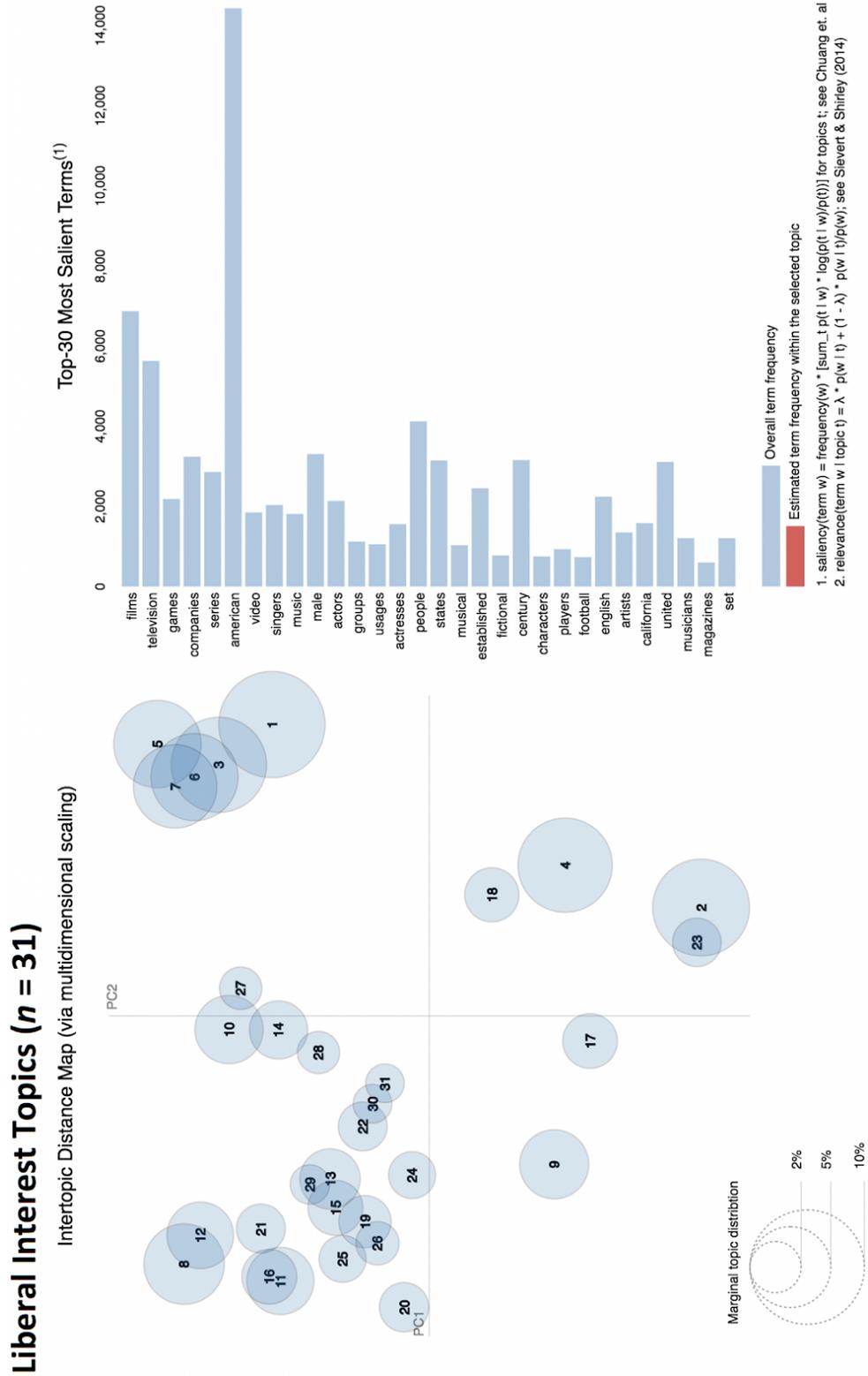